\documentclass[aps,prl,twocolumn,groupedaddress,floatfix,showpacs,keywords]{revtex4}

\usepackage{graphicx}
\usepackage{amssymb}

\begin{document}

\title{Granular impact and the critical packing state}

\author{Paul Umbanhowar}
\affiliation{Department of Mechanical Engineering,
Northwestern University, Evanston, IL 60208}

\author{Daniel I. Goldman}
\affiliation{School of Physics,
                Georgia Institute of Technology,
                Atlanta, GA 30332}
\date{\today}

\begin{abstract}
Impact dynamics during collisions of spheres with granular media reveal a pronounced and non-trivial dependence on volume fraction $\phi.$  Post impact crater morphology identifies the critical packing state $\phi_{cps},$ where sheared grains neither dilate nor consolidate, and indicates an associated change in spatial response. Current phenomenological models fail to capture the observed impact force for most $\phi$; only near $\phi_{cps}$ is force separable into additive terms linear in depth and quadratic in velocity.  At fixed depth the quadratic drag coefficient decreases (increases) with depth for $\phi<\phi_{cps}$ ($\phi>\phi_{cps}$).  At fixed low velocity, depth dependence of force shows a Janssen-type exponential response with a length scale that decreases with increasing $\phi$ and is nearly constant for $\phi > \phi_{cps}.$
\end{abstract}

\maketitle

Impact of objects into unconsolidated granular materials~\cite{jaeAnag} like sand is relevant in many settings (e.g.\ terminal ballistics of projectiles, hammered intruders, and ground-foot interaction~\cite{cheAumb09}) and is of scientific interest because the localized strain field around the impactor generates interaction between fluid and solid granular states. Unlike rapid granular flows which can be described by hydrodynamic-like equations~\cite{jenAric} (e.g.\ steady chute flow with no enduring contact networks), equivalent comprehensive and tested continuum descriptions for the mixed solid/fluid regime are lacking.

\begin{figure}
\begin{center}
\includegraphics[width=3.35in]{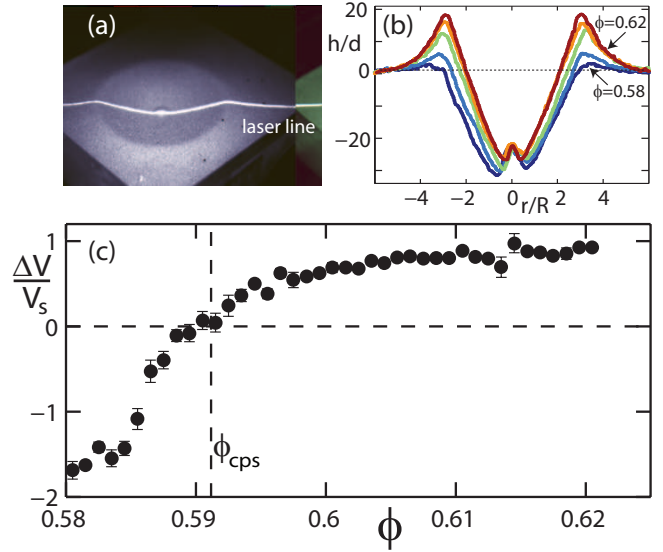}
\caption{(Color online) Influence of volume fraction on cratering at $v_0=257\pm3$~cm/s. (a) Impact crater at $\phi=0.61$ and laser line. (b) Surface displacement $h$ relative to grain diameter $d$ increases with volume fraction ($\phi=0.579,$ $0.589,$ $0.600,$ $0.610,$ and $0.622$). (c) Post-impact change in bed volume $\Delta V$ relative to sphere volume $V_s$ vs.\ $\phi$ is $0$ at $\phi_{cps} = 0.591,$ indicating the location of the critical packing state.}
\label{figureCrater}
\end{center}
\end{figure}

In the absence of governing equations and with a scarcity of direct force measurements, numerous phenomenological force models have been proposed over hundreds of years \cite{robins,allAmay57a,forAluk92,picAlar,debAwal04,lohAber04,houApen05,tsiAvol05,ambAkam05,wadAsen06,golAumb08} to explain the observed dependence of penetration depth, crater morphology, and collision duration on impact velocity and intruder and grain properties. To the best of our knowledge, all existing models assume that the granular resistance force $F$ can be separated into independent functions of position and velocity such that $F(z,v) = F_z(z) + F_v(v),$ where $z$ and $v$ are the projectile's depth below the initial free surface and velocity respectively. The depth dependent term $F_z$ has been modeled as constant~\cite{robins,debAwal04,tsiAvol05}, linear \cite{allAmay57a,debAwal04,golAumb08} and as a modified exponential~\cite{ambAkam05}. Velocity dependence also remains uncertain; it has typically been treated as an inertial drag, $F_v = \alpha v^2$ \cite{robins,allAmay57a,tsiAvol05,ambAkam05,golAumb08,katAdur07}, although linear \cite{debAwal04} and constant \cite{lohArau} forms have also been proposed.

No impact experiments or models have systematically examined the effect of volume fraction $\phi$, a parameter that largely determines the mechanical response of slowly sheared granular media \cite{neddermanbook}. $\phi$, the ratio of material volume (total mass divided by constituent solid density) to occupied volume, ranges between 0.55 and 0.64~\cite{jerAsch08} for dry non-cohesive granular media of slightly polydisperse and nominally spherical grains. In slow granular shearing, $\phi$ is a principle determinant of yield stress through changes in flow structure: for large $\phi,$ materials dilate and flow locally in shear bands while for small $\phi$ they consolidate and flow globally without shear bands~\cite{neddermanbook}. Only at the {\em critical packing state} \cite{schofieldbook} or CPS, which occurs at intermediate $\phi,$ is the volume fraction constant under shear.  Volume fraction effects are expected for penetration as well, and recent work examining slow constant velocity intrusion of a cylinder has observed signatures of a phase transition at $\phi \approx 0.60$~\cite{schroter07}.

Here we directly measure the time resolved impact force and post-impact crater profile to determine the influence of $\phi.$ In our experiments an $R=1.98$~cm radius steel sphere with total mass $m=270$~g is dropped with initial collision velocities $0 < v_0 < 350$~cm/s into the center of a $24\times24$~cm$^2$ cross-section box filled to a depth of $\approx 30$~cm with $d=300$~$\mu$m mean diameter glass spheres (density $\rho=2.52$~g/cm$^3$). An accelerometer on the sphere records vertical acceleration $a$ (see Ref.~\cite{golAumb08} for details) from which the non-dimensional penetration force is determined: $\tilde{F}=F/mg = a/g+1$, where $g$ is the acceleration due to gravity. The sphere's velocity is $v(t) = v_0 + \int_0^t a(t') dt'$ and its lowest point beneath the initial sand surface is $z(t)=\int_0^t v(t')dt',$ where $t=0$ corresponds to the time the bottom of the sphere first contacts the bed. Impactor and glass sphere dimensions ensure that our results are not influenced by finite size effects~\cite{neddermanbook} or interstitial air~\cite{pakAvan}.  Bed dimensions were chosen to eliminate boundary effects~\cite{segAber,nelAkat08}, but as we will suggest, the influence of horizontal walls appears to increase with increasing $\phi$.

States with  $0.57<\phi<0.63$ are generated by initially flowing air upward through the rigid distributor base of the bed to create a fully fluidized state, then decreasing air flow below fluidization onset and vibrating the container to reduce $\phi$ to the desired value~\cite{cheAumb09}.  Air flow and vibration are stopped during bed height measurement and after reaching the desired volume fraction: collisions occur in a quiescent bed. $\phi$ is measured to a precision of $0.001$ using an ultrasonic range finder to determine bed height. Additional x-ray absorption measurements confirmed that vertical variation in $\phi$ is small (less than 0.004 for $\Delta \phi=0.04$) as in previous observations, e.g.\ \cite{phiAbid02}. Surface profiles are measured using laser line profilometry~\cite{heiArer}, see Fig.~\ref{figureCrater}(a).

We first examine the impact crater (studied previously only at fixed $\phi$, e.g.~\cite{devAdeb07}) for constant $v_0$ and show that its shape and the amount of displaced material are sensitive to volume fraction. Figure~\ref{figureCrater}(b) indicates that at large $\phi$ the crater has a high rim and a small central peak (the remnant of a granular jet~\cite{lohAber04}), while at small $\phi$ the crater is deeper and has a lower rim and a larger central peak. To quantify these changes, we use the crater height profile $h(r)$, where $r$ is the radial distance from the crater center, to calculate the post-impact change in bed volume $\Delta V = 2\pi \int_0^\infty r h(r) dr - V_s$ vs.\ $\phi,$ where $V_s$ is the sphere volume (the sphere is always fully submerged post-impact). Figure~\ref{figureCrater}(c) shows that $\Delta V$ is negative at low $\phi$ and positive with smaller slope at high $\phi.$  $\Delta V$ can be interpreted as the volume of grains disturbed in the collision, $V_d,$ times the average change in volume fraction within $V_d.$ Since $V_d > 0,$ the transition in $\Delta V$ from negative (compaction) to positive (dilation) indicates that the critical packing state (no change in $\phi$ with shear) occurs at a volume fraction of $\phi_{cps}=0.591\pm 0.005$ for the glass beads used in this study. We quantify the proximity to CPS with $\Delta \phi=\phi-\phi_{cps}.$

\begin{figure}
\begin{center}
\includegraphics[width=3.35in]{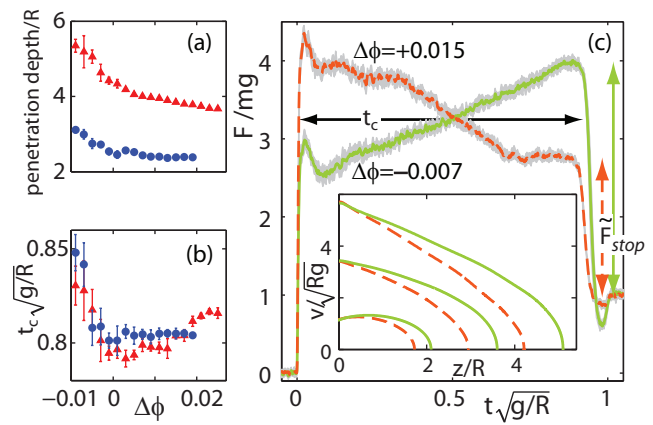}
\caption{(Color online) Effects of volume fraction on impact. (a) Penetration depth decreases with increasing $\Delta \phi$ while (b) collision duration is nearly unchanged [$v_0 = 134$~cm/s (blue circle) and $v_0 = 283$~cm/s (red triangle)]. In (a) and (b) changes in response occur near $\Delta \phi =0.$ (c)  Penetration force increases with time for $\Delta \phi=-0.007$ (green solid) but decreases for $\Delta \phi=+0.015$ (orange dashed) at $v_0=190$~cm/s. Curves are averages of 10 experiments; gray regions are $\pm 1\sigma$ and indicate the high degree of experimental repeatability. Inset: $z$$v$-plane trajectories differ with $\Delta \phi$ for the same $v_0.$}
\label{xfig}
\end{center}
\end{figure}

At the critical packing state the granular medium is effectively incompressible, suggesting the possibility of simpler impact dynamics in its vicinity and qualitative differences in dynamics between states with $\phi<\phi_{cps}$ and $\phi>\phi_{cps}$ due to $\phi$ dependent changes in flow as in Fig.~\ref{figureCrater}(c).  Beginning with kinematics, Fig.~\ref{xfig}(a) shows that the penetration depth at fixed $v_0$ decreases with increasing volume fraction as expected.  Penetration depth decreases sharply for $\Delta \phi < 0$ but more gradually for $\Delta \phi > 0$ with a total decrease of $\approx 30$\%. Collision duration $t_c$, although it varies little, is more sensitive to variation in $\phi$ for $\Delta \phi < 0,$ see Fig.~\ref{xfig}(b).  $\Delta \phi$ influences kinematics over the course of the collision as indicated by impactor trajectories in the $z$$v$-plane [inset of Fig.~\ref{xfig}(c)].

\begin{figure}
\begin{center}
\includegraphics[width=2.75in]{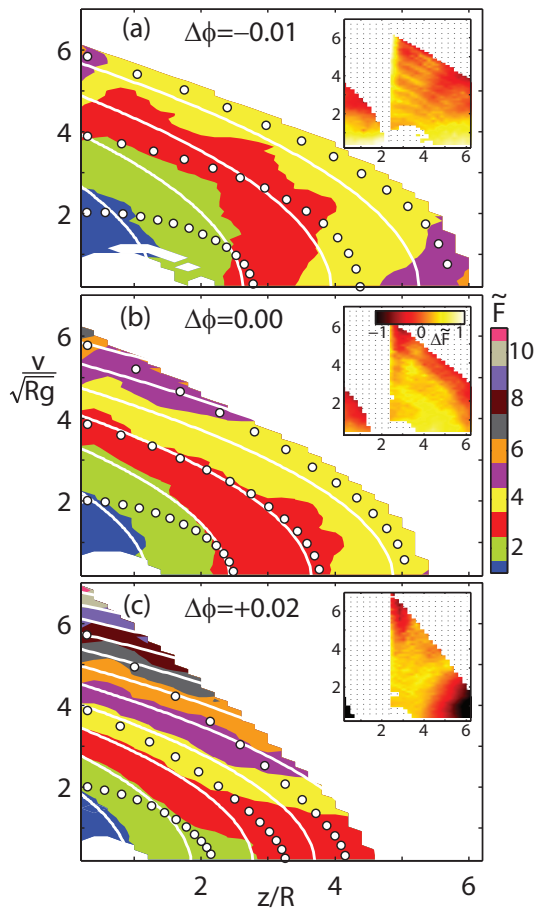}
\caption{(Color online) Isoforce contours (boundaries of colored regions) and impactor trajectories (circles, 6~ms sampling interval, $v_0/\sqrt{Rg}=2$, 4, 6) in the $z$$v$-plane differ (a) below, (b) at, and (c) above the critical packing state. At CPS, contours predicted by Eqn.~\ref{eq1} (white curves) best match experiment, and $\Delta \phi$ dependent deviations from Eqn.~\ref{eq1}, $\Delta \tilde{F} = \tilde{F} - \tilde{F}_u,$ are least localized. Eqn.~\ref{eq1} fit parameters: $k=0.62,$ 0.88, 1.21 and $\alpha=0.76,$ 0.82, 1.07 for $\Delta \phi=-0.01,$ 0.00, 0.02, respectively.}
\label{figGlobal}
\end{center}
\end{figure}

The $\phi$ dependence of kinematics originates in the impact force, which, as Fig.~\ref{xfig}(c) shows, differs above and below the critical packing. For example, at $v_0=190$~cm/s and below CPS ($\Delta \phi=-0.007$), $\tilde{F}$ increases with time from 2.5 to 4 times the weight of the intruder (like slow penetration in granular media~\cite{neddermanbook}), while above CPS ($\Delta \phi=+0.015$), $\tilde{F}$ decreases by about the same amount over the same time interval. In both, collision onset is marked by an initial jump in $\tilde{F},$ and collision termination is characterized by a sharp decrease in force of magnitude $\tilde{F}_{stop}$ \cite{golAumb08}. The average slope of $\tilde{F}(t)$ during collision decreases with increasing $\phi$ (and $v_0,$ see \cite{golAumb08}); the decrease is more rapid for $\Delta \phi < 0.$

To better characterize the dependence of impact force on volume fraction, depth, and velocity, we measured $\tilde{F}(t)$ for $-0.01<\Delta \phi<0.03$ varied in increments of 0.002 and for $0<v_0/\sqrt{Rg}<6$ varied in increments of $\approx5$~cm/s. For each sampled time in each collision we calculated the position $z(t)$ and velocity $v(t)$ of the impactor; note that $z$$v$-trajectories for distinct $\Delta \phi$ and $v_0$ do not intersect, see Figs.~\ref{xfig}(c) and \ref{figGlobal}. We then partitioned the $z$$v$-plane into \mbox{0.2~cm $\times$ 5~cm/s} regions and calculated the average force in each region to find $\tilde{F}(z,v,\Delta \phi).$ Figure~\ref{figGlobal} presents isoforce contours of $\tilde{F}(z,v,\Delta \phi)$ for $\Delta \phi$ below, at, and above CPS and reveals $\Delta\phi$ dependent changes in $\tilde{F}.$ For small $z,$ force increases more rapidly with $v$ as $\Delta \phi$ is increased, while for small $v,$ $\tilde{F}$ increases more slowly with depth for increasing $\Delta \phi.$  Impactor trajectories (circles in Fig.~\ref{figGlobal}) show that with increasing depth $\tilde{F}$ generally increases for $\Delta \phi < 0,$ decreases for $\Delta \phi >0,$ and changes least at $\Delta \phi=0.$

We compare our data to a model from a recent study by Katsuragi et al.\ \cite{katAdur07} of sphere impact into glass beads with initial $\phi=0.590.$ Using force data derived from high-speed imaging of impactor position vs.\ time, the authors proposed a ``unified'' force law of the form
\begin{equation}
\tilde{F}_u(z,v) = \frac{k}{R}z + \frac{\alpha}{Rg} v^2,
\label{eq1}
\end{equation}
where $k$ and $\alpha$ are dimensionless constants. We fit $k$ and $\alpha$ to $\tilde{F}(z,v)$ at each $\Delta\phi$ and found that Eqn.~1 best describes $\tilde{F}$ near CPS, see Fig.~\ref{figGlobal}. Below and above CPS, differences between isoforce contours of experimental data (boundaries between colored regions) and the model (white curves), $v/\sqrt{Rg} = \sqrt{(\tilde{F}_u - \frac{k}{R}z)/\alpha},$ are greatest at low $v.$ Insets in Fig.~\ref{figGlobal} plot differences between the experiment and model fits, $\Delta \tilde{F} = \tilde{F} - \tilde{F}_u,$ and show that for $\Delta \phi <0,$ the model underestimates $\tilde{F}$ at both small $z$ and small $v$, while for $\Delta \phi >0,$ it overestimates $\tilde{F}$ at large $z$ and small $v.$

\begin{figure}
\begin{center}
\includegraphics[width=3.35in]{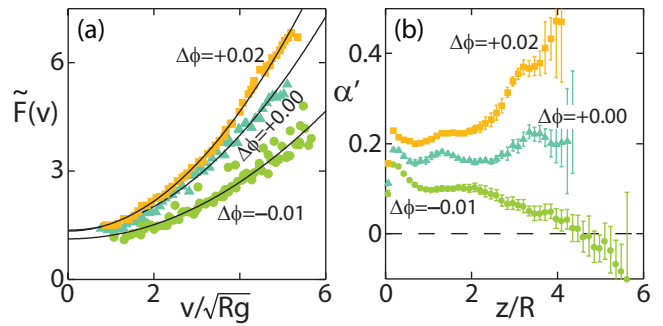}
\caption{(Color online) Variation of $\alpha'$ with depth. (a) $\tilde{F}$ and fits of $\tilde{F}\propto v^2$ (curves, see text) for $\Delta \phi$ below, at, and above CPS at fixed $z/R=1.1.$  (b) Variation of $\alpha'$ with depth depends on $\Delta \phi.$  Error bars increase as the range of measured $v$ decreases with $z$. Only near $\Delta\phi=0$ is $\alpha'$ depth independent. Error bars in (b) are 95\% confidence intervals.}
\label{veldep}
\end{center}
\end{figure}

Figure~\ref{figGlobal} suggests that a force model linear in depth and quadratic in velocity is insufficient away from CPS. Isolating the velocity dependent contribution by examining data at fixed depths (e.g.\ along vertical lines in Fig.~\ref{figGlobal}) using $\tilde{F} = C + \frac{\alpha'}{Rg} v^2,$ where $\alpha'$ and $C$ are constants free to vary with depth, shows this is the case (see Fig.~\ref{veldep}(a) for example fits at $z/R=1.1$). Figure~\ref{veldep}(b) shows that $\alpha'$ decreases with $z$ for $\Delta \phi < 0$ and increases with $z$ for $\Delta \phi > 0,$ which rules out a separable penetration force with a purely $v^2$ velocity dependence.  Only near CPS is $\alpha'$ independent of depth as in most granular impact-force models~\cite{robins,allAmay57a,forAluk92,picAlar,debAwal04,lohAber04,houApen05,tsiAvol05,ambAkam05,wadAsen06,golAumb08}.  We note, however, that by adding a term linear in velocity, i.e.\ $\tilde{F} = C + \frac{\beta}{\sqrt{Rg}}v+\frac{\alpha'}{Rg} v^2,$ $\alpha'$ becomes always positive and nearly independent of depth at all $\phi$.

\begin{figure}
\begin{center}
\includegraphics[width=3.35in]{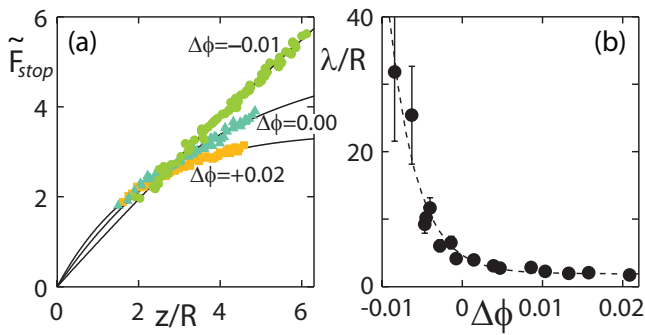}
\caption{(Color online) $\Delta \phi$ driven changes in grain force at the end of collision.  (a) $\tilde{F}_{stop}$ is linear in depth at the lowest $\Delta \phi$ but becomes increasingly sub-linear as $\Delta \phi$ increases; curves are fits of $\tilde{F}_{stop}$ vs.\ $z/R$ to a two parameter model similar to Janssen's Law (see text). (b) Characteristic length $\lambda$ from model decreases rapidly with $\Delta \phi$ below CPS but is nearly constant above it (dashed curve is an exponential fit).  Error bars in (b) are 95\% confidence intervals.}
\label{figureLambda}
\end{center}
\end{figure}

The effects of proximity to CPS are also evident at low velocity where depth dependent frictional forces are expected to dominate. $\tilde{F}_{stop}$ (see Fig.~\ref{xfig}) characterizes the low velocity response at the end of collision and is plotted in Fig.~\ref{figureLambda}(a) vs.\ depth for representative $\Delta \phi$. For the smallest $\Delta \phi,$ $\tilde{F}_{stop}$ increases linearly with depth while at larger $\Delta \phi$ the dependence is sub-linear. The force in the depth dominated regime for all $\Delta \phi$ is well modeled by a single Janssen-like function~\cite{janssen1895} $\tilde{F}_{stop} = k'\left ( 1 - \mathrm{e}^{-z/\lambda} \right)$ [black curves in Fig.~\ref{figureLambda}(a)], where $\lambda$ is a characteristic length and $k'$ a constant. $\lambda$ decreases rapidly from $\approx 30R$ to $\approx 2R$ as $\Delta \phi$ approaches zero from below, see Fig.~\ref{figureLambda}(b). For $\Delta \phi>0$, $\lambda$ is nearly constant.  In the limit of large $\lambda$ (low $\Delta \phi$), the model gives a linear response $\tilde{F}_{stop}\approx \frac{kz}{\lambda}$ with $k=k'/\lambda$ as in Eqn.~\ref{eq1}, while in the opposite extreme, $\tilde{F}_{stop}$ is constant as in Ref.~\cite{tsiAvol05}.  To determine if $\lambda$ is also dependent on container size, we reduced the width of the fluidized bed from 24~cm to 12~cm and found that $\lambda$ decreased by $\approx 2/3$ at all $\Delta\phi.$  Our findings suggest that as $\phi$ is increased, grains exert increasingly larger forces on the sidewalls, and presumably also on the intruder, at shallower depths which ultimately reduce the gravitational forces on the grains leading to a net reduction in the depth dependent component of the penetration force.  Whether or not $\tilde{F}_{stop}$ is ultimately linear in $z$ at large $\Delta\phi$ in an unbounded container is an open question.

By varying the volume fraction, we have shown that the dynamics of impact in granular media are richer than previously thought and that existing separable models of granular impact linear in depth and quadratic in velocity do not capture all the details, likely due to changes in flow and the influence of boundaries associated with compaction and dilation. Dynamics in spatially extended systems based on ODE's are often incomplete; in the case of granular impact, they fail to explain why and how heuristic parameters such as $k$ and $\alpha$ change with depth, velocity, and $\phi.$ To completely characterize granular impact it is likely that the full spatio-temporal response of the granular ensemble is needed; continuum approaches capturing the transition from static to flowing regimes appear promising, see for example~\cite{volAtsi03}.

\begin{acknowledgments}
We thank Harry Swinney and Matthias Schr\"{o}ter for helpful discussions and Chen Li for preliminary data.  This work was supported by the Burroughs Wellcome Fund and the Army Research Laboratory (ARL) Micro Autonomous Systems and Technology (MAST) Collaborative Technology Alliance (CTA) under cooperative agreement number W911NF-08-2-0004.
\end{acknowledgments}


\end{document}